\documentclass[journal=jacsat,manuscript=article]{achemso}

\usepackage[version=3]{mhchem} 
\usepackage[T1]{fontenc}       
\usepackage{graphicx}
\usepackage{caption}
\usepackage{subcaption}

\author{Iman Esmaeil Zadeh\footnote[0]{$\ast$ These authors contributed to this work equally}}
\affiliation[TU delft]
{Kavli Institute of Nanoscience Delft, Delft University of Technology, Delft, the Netherlands.}

\alsoaffiliation[SQ]
{Single Quantum B.V., Delft, The Netherlands.}

\email{i.esmaeilzadeh@tudelft.nl}

\author{Johannes W. N. Los\footnotemark[0]}
\affiliation[SQ]
{Single Quantum B.V., Delft, The Netherlands.}

\author{Ronan B. M. Gourgues}
\affiliation[SQ]
{Single Quantum B.V., Delft, The Netherlands.}

\author{Gabriele Bulgarini}
\affiliation[SQ]
{Single Quantum B.V., Delft, The Netherlands.}

\author{Sergiy M. Dobrovolskiy}
\affiliation[SQ]
{Single Quantum B.V., Delft, The Netherlands.}

\author{Val Zwiller}
\affiliation[KTH]
{Department of Applied Physics, Royal Institute of Technology (KTH), Stockholm, Sweden}
\alsoaffiliation[TU delft]
{Kavli Institute of Nanoscience Delft, Delft University of Technology, Delft, the Netherlands.}
\alsoaffiliation[SQ]
{Single Quantum B.V., Delft, The Netherlands.}

\author{Sander N. Dorenbos}
\affiliation[SQ]
{Single Quantum B.V., Delft, The Netherlands.}

\title[An \textsf{achemso} demo]
  {A single-photon detector with high efficiency and sub-10\,ps time resolution}


\begin{document}


\begin{abstract}
The observation of fast physical dynamics using optical techniques currently relies on indirect methods, such as pump-probe measurements. One reason for this is the lack of an efficient detector with high time resolution. Single-photon detectors with high efficiency and ultra-high time resolution serve as the best candidates for replacing such indirect methods. We engineer the nano-structure of Superconducting Nanowire Single-Photon Detectors (SNSPDs) to achieve a time resolution better than 10\,ps and at the same time a high efficiency ($>86\%$). Furthermore, at the limit of multiphoton excitation, using an improved readout technique, we reach an unprecedented time resolution of <3\,ps. These findings set a new upper limit for the intrinsic time resolution of SNSPDs and open up new possibilities for direct observation of fast phenomena in different fields of science.    
\end{abstract}

\section{Introduction}

Superconducting Nanowire Single-Photon Detectors (SNSPDs) are successfully used in many applications ranging from CMOS testing \cite{Zhang:2003}, biomedical imaging \cite{Peer:2007}, laser ranging \cite{McCarthy:2013}, and quantum communication \cite{Yin:2013,Vallone:2016}. These detectors have proven unparalleled performance, being the only technology that reaches ultra-high efficiency, time resolution, and count-rate at the same time \cite{Zadeh:2017}. Yet there are many fast phenomena in chemistry, biology, physics \cite{McCarthy:2013}, where detectors with superior timing jitter are required.  Moreover, a practical device, for the mentioned applications, also requires good performance in terms of efficiency.  Here, we show a device combining unprecedented time resolution in conjunction with high efficiency

\section{Device design and fabrication}

Similar to \cite{Zadeh:2017}, we fabricate our detectors from sputtered NbTiN films on top of a $\lambda$/4 SiO2 cavity, finished by a gold mirror. To achieve the highest possible absorption in the superconducting layer and also increase the critical current of the device, enhancing signal to noise ratio and thus the jitter, thicker films are required \cite{Zadeh:2017}. On the other hand, increasing the film thickness enlarges the superconducting bandgap of the film. Therefore, for such films, an absorbed photon will lead to fewer excited quasiparticles that  can form the resistive region across the nanowire. This means that for these films, it is more challenging to reach saturation of the internal efficiency. To make the detectors reach full internal efficiency, we perfected the fabrication of the nanowire meander to achieve the highest uniformity along the nanowire length. This allowed us to saturate detectors made out of a 9\,nm thick film (top section of the film contains 1-1.5\,nm oxide \cite{Cheng:2016}, preventing it from further oxidation) with 100\,nm/120\,nm wide nanowire (filling factors of 0.5/0.6), at wavelengths below 850nm. To make high critical current, good risetime, and high efficiency detectors at 1550\,nm, we fabricated detectors from a 8.5\,nm thick film with $\sim$ 45\,nm wide nanowires (and filling factor of 0.4). 
Figure 1 shows examples of devices fabricated for 800-900\,nm and 1550\,nm.

\begin{figure}
	\centering
	\includegraphics[scale=0.4]{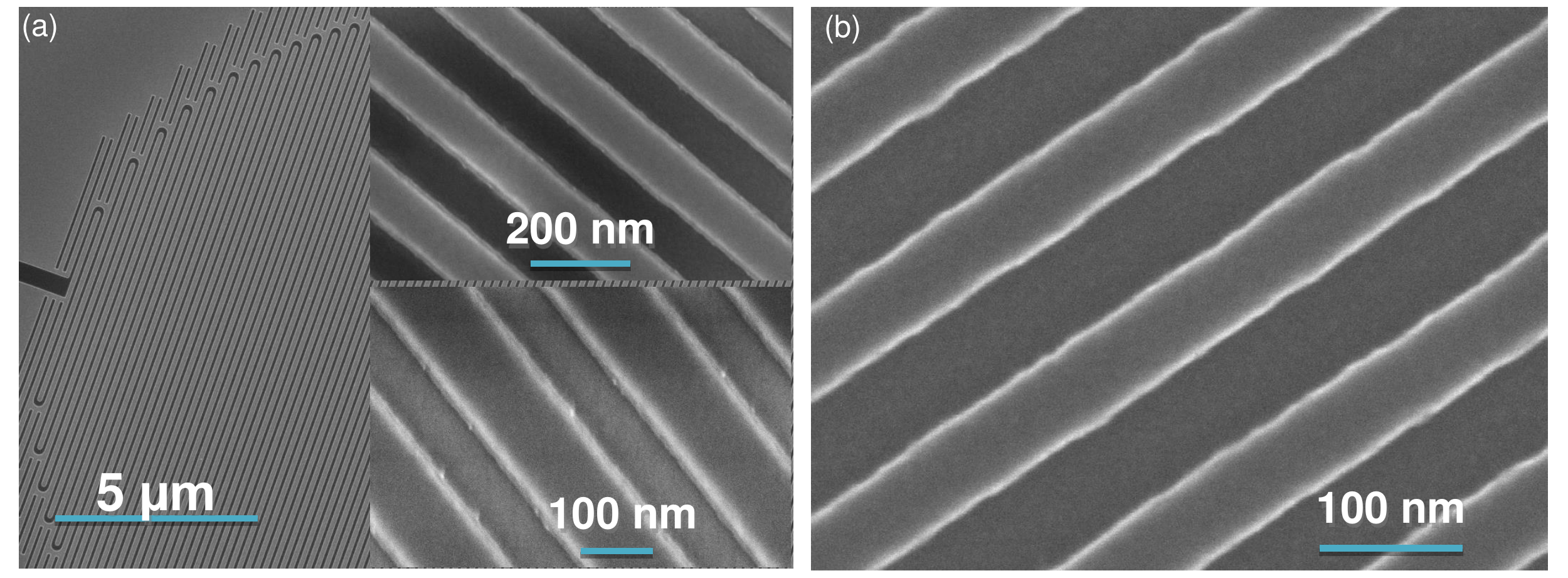}
	\caption{Scanning electron microscope (SEM) images of some fabricated devices. (a) An example of devices fabricated for 800-900\,nm. Top right inset shows device with filling factor of $\sim$0.55, and bottom inset is an example of a device with filling factor of 0.6. (b) An example of a detector fabricated for 1550\,nm. Usual filling factors are  0.4-0.45.}
	\label{fig:fig1}
\end{figure}

\section{Measurements}

\begin{figure}
	\centering
	\includegraphics[scale=0.4]{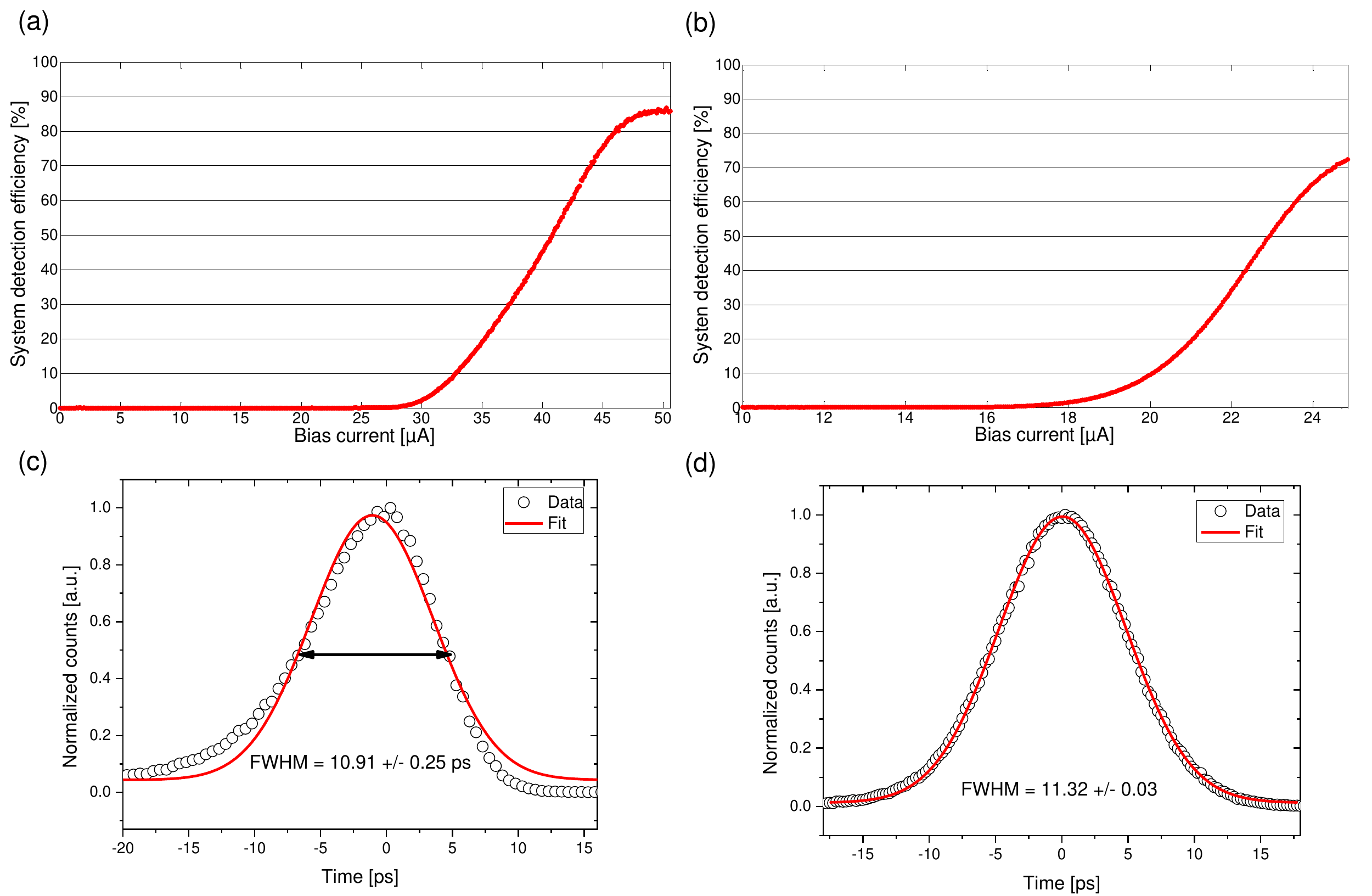}
	\caption{(a) Efficiency versus bias current for a detector at 787\,nm. (b) Efficiency versus bias current for another detector which is optimized for the wavelength of 1550\,nm. (c) Jitter measurement for the detector shown in (a). (d) Jitter measurement for the detector shown in (b)}
	\label{fig:fig2}
\end{figure}

We carefully test the performance of our detectors in a standard setup similar to \cite{Zadeh:2017}.
Figure 2a and Figure 2b present the efficiency curve versus bias current for two detectors optimized for wavelengths of 787\,nm and 1550\,nm, reaching $>86\%$, and $>72\%$, respectively. Both detectors have high critical current and fast risetime. After careful characterization of the efficiency, we measure the detectors timing jitter. The cryogenic amplifiers that are used for measuring the two types of detectors (787\,nm and 1550\,nm) are slightly different in terms of their gain and bandwidth. This choice is due to their different rise-time and critical current. The amplifiers information can be found in the supplementary information. The jitter measurements for 787\,nm and 1550\,nm detectors are shown in Figure\,~\ref{fig:fig2}c and Figure\,~\ref{fig:fig2}d, respectively. These detectors have the lowest reported SNSPD time jitter to date. 

For the detector shown Figure\,~\ref{fig:fig2}b and d, we measured the dependence of the jitter on the bias current, the result is shown on Figure\,~\ref{fig:fig3}a. When room temperature amplifiers are used, the jitter decreases almost asymptotically with the increase of the bias current. For the case of cryogenic amplifiers, however, the value for jitter seems to get  close to saturation.

The count-rate of the detector can also influence the jitter and  found out that rather than only detector countrate, the ratio of the detector countrate to the laser rep-rate highly influences the shape of the distribution. We attenuated pulses from a 50\,MHz pulsed laser to get different detection rates and carried out photon correlations at each point, the results of such measurements for the same device as Figure\,~\ref{fig:fig2}a, are presented in Figure\,~\ref{fig:fig3}b. It can be observed that at low count-rates (below 0.1 $\times$ laser repetition-rate), SNSPD time jitter exhibit a typical single Gaussian distribution. However, as the count-rate increases, other peaks appear in the distribution. The observation of these peaks can be explained by the fact that when the recovery time of the bias current of SNSPD is longer than $\frac{1}{laser\,rep.\,rate}$, subsequent detection events take place with an under-biased detector, while detection events which are temporally further away would experience detection with full bias current.  

\section{Multi-photon excitation and sub-3\,ps jitter}

\begin{figure}
	\centering
	\includegraphics[scale=0.4]{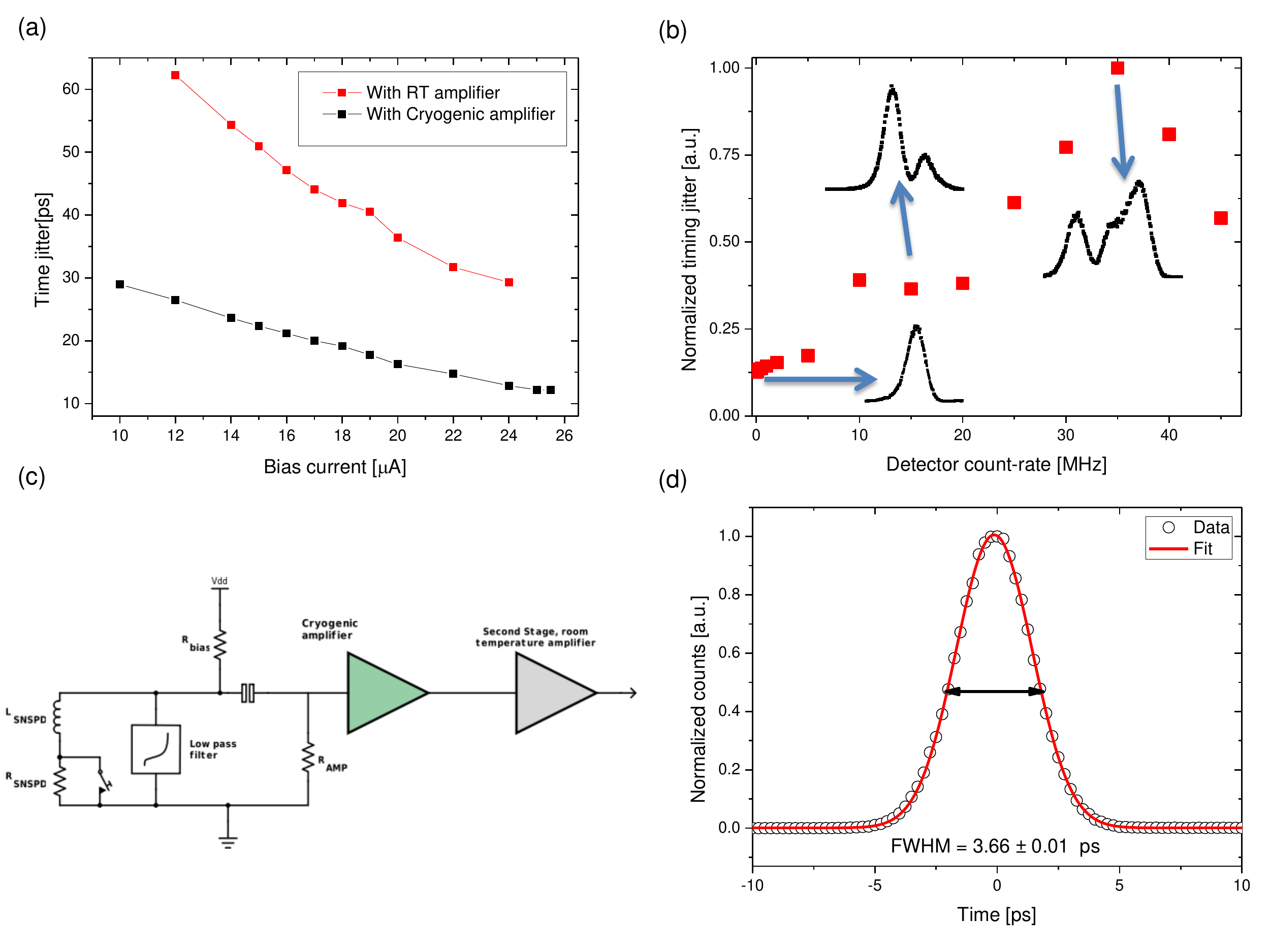}
	\caption{(a) Dependence of jitter on bias current for the cases of room temperature and cryogenic amplification. (b) Jitter vs count-rate. At higher count-rates the jitter of the detector is not a single Gaussian. This is because a percentage of the detection pulses (the percentage depends on the count-rate and electrical dead-time of the detector) interfere with each another and modify the amplitude, slope, and offset. The interfering pulses cause extra peaks in the distribution of the jitter. (c) Schematic of low-pass filter readout of SNSPD for multi-photon excitation. (d) At the limit of multiphoton excitation, a jitter of 3.66\,ps($<$3\,ps decoupled) can be achieved. } 
	\label{fig:fig3}
\end{figure}

Normally SNSPDs cannot be operated under high excitation powers (typical SNSPD exication powers are in the femtowatt range giving countrates below 1\,MHz) because they latch into the normal state (they become normal with no recovery to the superconducting state). By providing a low frequency path for the SNSPD current to the ground, it is possible to operate the detector at higher optical powers. Moreover, by delivering high frequency components of the pulse to the amplifier, a good jitter can be achieved. We use a low-pass filter, as shown in Figure\,~\ref{fig:fig3}c, in parallel with the detector. 

The low-pass filters made it possible to excite the detector with powers in the range of 1-10\,nW. With higher excitation powers it was possible to reach countrates that matched the repetition rate 50 MHz of the laser we used. Therefore every single optical pulse from the laser could be filled with photons so that the optical pulse length did not contribute to the time jitter of our SNSPDs. Moreover, at these high powers, any intrinsic/geometric contribution to the jitter is averaged out as only the fastest events contribute to the distribution. Therefore, the only cause of jitter was the combination of the electrical noise and the finite amplifier bandwidth (see supplementary information). The results of time jitter measurement with multi-photon excitation is shown in Figure\,~\ref{fig:fig3}d, demonstrating a remarkable high time-resolution of 3.66\,ps ($<$3\,ps after decoupling the contributions from the correlator and the photodiode, see supplementary).

\begin{acknowledgement}
	Ronan B.M Gourgues acknowledges support by the European Commission via the Marie-Sklodowska Curie action Phonsi (H2020-MSCA-ITN-642656). This research was supported by Single Quantum B.V.. 
\end{acknowledgement}

\bibliography{LJHE}

\end{document}